\documentclass[12pt]{article}
\usepackage{amssymb}
\usepackage{amsfonts}
\usepackage{amsmath}
\usepackage{graphicx}
\usepackage{latexsym}

\newcommand{\bb}{\begin{eqnarray}}
\newcommand{\ee}{\end{eqnarray}}
\newcommand{\beq}{\begin{equation}}
\newcommand{\eeq}{\end{equation}}
\newcommand{\ba}{\begin{array}}
\newcommand{\ea}{\end{array}}

\title{Multicomponent integrable wave equations I. Darboux--Dressing Transformation}

\author{Antonio Degasperis$^{(+)}$, Sara Lombardo$^{(\times)}$\\
\\
$^{(+)}$ Dipartimento di Fisica, Universit\`a di Roma 
``La Sapienza'' \\
and Istituto Nazionale di Fisica Nucleare, Sezione di 
Roma,\\ 
Rome, Italy. E-mail: antonio.degasperis@roma1.infn.it\\ 
$^{(\times)}$ Department of Mathematics, Vrije Universiteit, 
Amsterdam, NL.\\ E-mail: sara@few.vu.nl}

\date{}

\begin{document}

\maketitle
\begin{abstract}
The Darboux--Dressing Transformations are applied to the Lax pair
associated to systems of coupled nonlinear wave equations
in the case of boundary values which are appropriate to both ``bright'' and ``dark''
soliton solutions. The general formalism is set up and the relevant
equations are explicitly solved. Several instances of multicomponent wave equations of applicative interest, such as vector nonlinear Schr\"{o}dinger--type equations and three resonant wave equations are considered.

\vspace{1cm}

\end{abstract}

\newpage

\section{Introduction}
The theory of solitons originated long time ago with the discovery of the
integrability of the Korteweg-de Vries (KdV) equation and
with the method of the spectral transform to solve it \cite{GGKM67}. Many other
soliton equations, with various degrees of novel mathematical
features and applicative interests (see, for instance, \cite{CD82} and 
\cite{AC91}), were then found by extending and generalizing the method of the spectral transform. Among these integrable equations, the most
notable one is the nonlinear Schr\"{o}dinger 
(NLS) equation (subscripted variables denote partial differentiation and $x$ and $t$ are the independent variables):
\begin{equation}
u_t=i\gamma(u_{xx}-2s|u|^2u) ,\quad u=u(x,t),\quad s=\pm 1 \,,  
\label{nls}
\end{equation}
where $\gamma$ is the real dispersion parameter. The great importance of this equation is due to  its integrability and universality \cite{Calogero91} which are related consequences of a perturbative multiscale analysis of (a large class of) dispersive nonlinear wave equations (see also \cite{DMS1997}).
 
The key property of integrable evolution Partial Differential Equations (PDEs)
is that they express the condition that two linear differential
equations (Lax pair), both for the same unknown function and whose
coefficients depend also on a complex (spectral) parameter, are compatible
with each other. The compatibility of these two equations is  a local
condition and it provides local properties of the associated integrable
PDE, in particular local conservation laws (continuity equations) and
Hamiltonian structures. The construction of solutions of integrable
nonlinear evolution PDEs is, however, a different matter. In fact, in
addition to the initial value, one has to specify the domain of the space
variable
$x$ together with the values which the solution (and/or its
$x$--derivatives) takes on the domain boundary. Once the (appropriate)
boundary values are fixed, then one should make use of the Lax pair to solve the Cauchy initial
value problem or, more modestly, to construct  special
solutions. In both cases, various methods of solution have been devised
depending on the boundary conditions imposed on the solution, and several
problems, for certain initial--boundary values, remain open. 

In this paper we give the explicit construction of special solutions based on Darboux--Dressing Transformations  by means of algebraic and local operations only, with no need to referring to specific boundary values. The  construction of soliton solutions, with their appropriate boundary conditions,  will be considered in a subsequent paper for the case in which the space domain is the
entire real axis, $-\infty<x<+\infty$. Historically, the 
initial--value problem for the KdV equation in this space domain with
vanishing values at the boundary, $x=\pm \infty$, was the first to be
solved. Then, by the same spectral technique but applied to a different
Lax pair, Zakharov and Shabat were able to solve the
initial--value problem for the NLS equation~(\ref{nls}) on the entire
real $x$--axis and for both vanishing \cite{ZS71} and 
nonvanishing \cite{ZS73}  values at
the boundary $x=\pm\infty$. The possibility to solve these two
distinctive cases turned out to be quite relevant in nonlinear optics
where the soliton behaviour in both cases has been observed. If
the boundary values are vanishing, soliton solutions of (\ref{nls})
exist for $s=-1$, while if the boundary values are nonvanishing (and
appropriately prescribed) stable soliton solutions of
~(\ref{nls}) exist for $s=+1$, while for $s=-1$ solitons exist but are
unstable against small perturbations \cite{McL93}. Solitons are
usually referred to as ``bright'' solitons in the first case (vanishing boundary values) 
as they are
light-pulses in a dark background, and as ``dark'' solitons in the second
case (non vanishing boundary values) as they are pulses of darkness in 
a light background. In the latter
case a further distinction is made between ``black'' and ``grey'' solitons.
  For a guide to the vast literature on solitons see, for instance,
\cite{Degasperis98}.

The propagation of pulses in nonlinear media is more generally modelled
by multicomponent fields, and this requires to generalize the NLS equation
(\ref{nls}) to a system of coupled nonlinear
Schr\"{o}dinger equations. The coupling between the field components,
which may represent different polarization amplitudes or fields with
different (resonating) frequencies, as dictated by physical
contexts, may or may not lead to integrable nonlinear equations.
Though physical models may have terms which make the propagation equation
nonintegrable, still the investigation of sufficiently close integrable
equations, because of our good analytical control of them, is certainly
worth and valuable also in an applicative context. Integrable systems of
coupled NLS equations, termed vector nonlinear Schr\"{o}dinger (VNLS)
equations as the dependent variable $\mathbf{u}(x,t)$ is a
$D$--dimensional vector,
$\mathbf{u}=(u^{(1)} ,\cdots ,u^{(D)})$, are known with various
coupling terms. For instance, for  $D=2$ the system
\begin{equation}\label{manakov}
\begin{array}{ll}
u^{(1)}_t=i\gamma [u^{(1)}_{xx}-2s(|u^{(1)}|^2+|u^{(2)}|^2)u^{(1)}]\,,&  \\ 
u^{(2)}_t=i\gamma [u^{(2)}_{xx}-2s(|u^{(1)}|^2+|u^{(2)}|^2)u^{(2)}]\,, & s=\pm 1 \,,
\end{array}
\end{equation}
was first introduced by Manakov \cite{Manakov73}.
Similarly to  the scalar 
NLS equation (\ref{nls}), also this system possesses bright--solitons (i.e. with
vanishing boundary values) for $s=-1$ and (stable) dark--solitons
(i.e. with nonovanishing boundary values) for $s=+1$. This system (\ref{manakov}) remains integrable even if generalized to
\begin{equation}\label{s1s2}
\begin{array}{ll}
u^{(1)}_t=i\gamma [u^{(1)}_{xx}-2(s_1|u^{(1)}|^2+s_2|u^{(2)}|^2)u^{(1)}]
\,,& \\ 
u^{(2)}_t=i\gamma [u^{(2)}_{xx}-2(s_1|u^{(1)}|^2+s_2|u^{(2)}|^2)u^{(2)}]
\,, & s_j=\pm 1, ~\;j=1,2 \,; 
\end{array}
\end{equation}
in fact, only three systems have been found integrable for two
NLS equations coupled by square modulo terms of this form, namely
equation (\ref{s1s2}) with $s_1=s_2=\pm 1$ and
$s_1=-s_2=1$ \cite{ZS}.

Different ways of coupling two or more  Schr\"{o}dinger
equations in a nonlinear way which preserves integrability have been
discovered. Some of them are included in the class of wave equations considered here and are displayed in section 2. Additional systems of wave equations which do not posses the 
Schr\"{o}dinger--type second order dispersion term since they are only first order in the $x$--derivatives, 
and therefore they may or may not model dispersive waves, are also treated here. The most notable example of such systems is the one describing the coupling of three waves at resonance (see section \ref{sec:mother}), a model which applies to several physical contexts such as optics, fluid dynamics and plasma physics.

All  systems of coupled Partial Differential Equations (PDEs) discussed in this paper  are particular 
reductions of a general matrix evolution equation, namely the matrix partial differential equation (\ref{boomeron}).  
The integrability of this ``mother'' equation, which is guaranteed
by the associated Lax pair, entails the possibility to solve the initial
value problem by the spectral transform (or inverse scattering) method.
However, the task of setting up the
formalism and finding out the relevant integral equation 
has not been carried out completely. Indeed, this task is
certainly easier if one assumes that the solution vanishes at the
boundary, while in the case of nonvanishing boundary values it is rather involved because of the  many branch points which occur in the complex plane of the spectral variable.
Even in the case of the $D$--dimensional VNLS (\ref{Dvector}), which is  a simple reduction of the system (\ref{boomeron}), the spectral method in the case of nonvanishing boundary values, well understood for the scalar NLS equation (\ref{nls}) \cite{FT86}, has not been taken to a state which is amenable to easy access in applications (the case $D=2$ has been recently considered in \cite{abp} and partial results have been obtained in \cite{eei}).  On the other hand, in the VNLS case, the dark vector soliton phenomenology has actracted considerable attention among nonlinear optics scholars, and explicit soliton expressions have been derived by direct methods (for instance Hirota method) \cite{KT}--\cite{N}, rather than by using the Lax pair.

The aim of this paper is  to provide a general method of
explicit construction of  soliton solutions of the mother equation (\ref{boomeron}), and therefore of all the various wave propagation equations which obtain from it by reductions. This method makes use of the Lax pair and goes via the Darboux-Dressing
Transformations (DDT) \cite{DDT1}--\cite{DDT4} in a standard way; see also \cite{DL}. However, the nonvanishing of the boundary values at
infinity in the variable $x$ introduces novel features with respect to
the more usual context of vanishing boundary values. In fact, the novelty here regards more the case of
dark solitons, rather than that of bright solitons. 

In the next section \ref{sec:mother} we introduce the general formalism and display some of the reduced  equations which we deem representative of potential applicative models. Some of these wave equations  follow from a generalization which has been recently introduced in \cite{CD2004} and \cite{CD2006} with the purpose of modelling solitons which behave as \emph{boomerons} (or \emph{trappons}) (see also \cite{CD3wave} and \cite{DCBW2006}).  Section \ref{sec:DDT} is devoted to the DDT technique in its general setting. Section \ref{sec:DDTcompl},  deals with DDTs characterized by a  complex (non real) pole, while section \ref{sec:DDTreal} deals with DDTs with a real pole, these two cases being appropriate to cope with different boundary values. In the last
section \ref{sec:conclusions}  we add comments and remarks.

%% Section 2
\section{General formalism and multicomponent wave equations}\label{sec:mother}
All systems of coupled PDEs considered in this paper are special
(reduced) cases of the following matrix PDE:
\setlength\arraycolsep{2pt}
\begin{equation}\label{boomeron}
\begin{array}{lll}
Q_t &=&[C^{(0)},Q]+\sigma [C^{(1)},Q_x]-\sigma
\{Q,W\}-i\gamma \sigma(Q_{xx}-2Q^3)\,,\\
W_x &= &[C^{(1)},Q^2]\,, 
\end{array}
\end{equation}
%\begin{equation}\label{mother}
%Q_t=-i\gamma \sigma (Q_{xx}-2Q^3)\,,\quad Q=Q(x,t)\,,
%\end{equation}
where the dependent variables $Q=Q(x,t)$ and $W=W(x,t)$ are $(N^{(+)}+N^{(-)})\times
(N^{(+)}+N^{(-)})$ block matrices of the form
\begin{equation}\label{qwblock}
Q=\left( \begin{array}{cc} \mathbf{0}_{N^{(+)}\times N^{(+)}} & Q^{(+)}
\\ Q^{(-)} & \mathbf{0}_{N^{(-)}\times N^{(-)}} \end{array} \right) \,,\quad
W=\left( \begin{array}{cc}  W^{(+)} & \mathbf{0}_{N^{(+)}\times N^{(-)}}\\ 
\mathbf{0}_{N^{(-)}\times N^{(+)}} & W^{(-)} \end{array} \right) \,.
\end{equation}
In self-evident notation, the diagonal entries $W^{(+)}$, $W^{(-)}$ are square matrices
of dimension $N^{(+)}\times N^{(+)}$ and $N^{(-)}\times N^{(-)}$, while the two off-diagonal rectangular blocks
$Q^{(+)}$, respectively $Q^{(-)}$, are
$N^{(+)}\times N^{(-)}$, respectively $N^{(-)}\times N^{(+)}$,
matrices, $N^{(+)}$ and $N^{(-)}$ being arbitrary, positive
integers. Here, and in the following, the constant diagonal
matrix $\sigma$ in (\ref{boomeron}) is, in self-evident notation,
\begin{equation}\label{sigma} 
\sigma=\left( \begin{array}{cc} \mathbf{1}_{N^{(+)}\times N^{(+)}} & 
\mathbf{0}_{N^{(+)}\times N^{(-)}} \\ \mathbf{0}_{N^{(-)}\times
N^{(+)}} & -\mathbf{1}_{N^{(-)}\times N^{(-)}} \end{array} \right)\,,
\end{equation}
while $C^{(0)}$, $C^{(1)}$ are arbitrary constant block--diagonal
matrices,
\begin{equation}\label{Cmatrices}
C^{(j)}=
\begin{pmatrix}
C^{(j)(+)} & \mathbf{0}_{N^{(+)}\times N^{(-)}} \\ 
\mathbf{0}_{N^{(-)}\times N^{(+)}} & C^{(j)(-)}
\end{pmatrix}\,,\quad j=0,1, 
\end{equation}
 and the block--diagonal matrix $W=W(x,t)$ is an auxiliary
dependent variable. As usual $[A,B]$ and $\{A,B\}$ are the
commutator $AB-BA$ and, respectively, the anticommutator $AB+BA$. 
The constant coefficient $\gamma$ is the real dispersion parameter, $\gamma =\gamma^{\ast}$.
The systems of coupled NLS equations which are obtained by reduction of
equation (\ref{boomeron}) have been considered in \cite{CD2004}, the special case with $\gamma=0$ having been extensively reported in \cite{CD2006} and \cite{CD3wave}, where it is shown
that their single soliton solutions feature boomeronic and
trapponic behaviours. 
It is worth noticing that setting $C^{(1)}=c\sigma$, and therefore $W=0$, the linear terms
$[C^{(0)},Q]$ and $\sigma[C^{(1)},Q_x]$ can be both transformed away by the obvious
transformation $Q(x,t)\rightarrow \hat{Q}(x,t)=\exp(-C^{(0)}t)
Q(x-2ct,t)\exp(C^{(0)}t)$ and one obtains the matrix evolution
equation 
\begin{equation}\label{mother}
Q_t=-i\gamma \sigma (Q_{xx}-2Q^3)\,,
\end{equation}
that is, the standard matrix version of the NLS type equation. However, whenever $C^{(1)}$ is a generic full block--diagonal matrix (see (\ref{Cmatrices})), equation (\ref{boomeron}) is the most general second order differential equation which
genuinely generalizes the standard equation (\ref{mother}). Indeed, it allows to treat the somewhat simpler class of evolution equations corresponding to the choice  
$\gamma=0$, the canonical form of which only features first derivatives with respect to the space variable $x$ (rather than second derivatives as is the
case of NLS type equations). The simplest system of
equations of this type can in fact be reformulated so as to coincide with
the standard equations describing the resonant interaction of 3 waves (see below) with the remarkable consequence that its solutions may feature the boomeronic or
trapponic behavior, as recently reported in \cite{CD3wave} and \cite{DCBW2006}.

% Lax pair
\subsection{Lax Pair and reductions}\label{sec:lp}
The matrix equation (\ref{boomeron}) is the compatibility condition for
the Lax pair
\begin{equation}\label{lax} 
\psi_x=X\psi\,, \quad\psi_t=T\psi\,, 
\end{equation}
where $\psi$, $X$ and $T$ are $(N^{(+)}+N^{(-)})\times
(N^{(+)}+N^{(-)})$ square matrices, $\psi=\psi(x,t,k)$ being a common
solution of the two linear ordinary differential matrix equations
(\ref{lax}) while $X=X(x,t,k)$ and $T=T(x,t,k)$ depend on the coordinate $x$
, the time
$t$ and the complex spectral parameter $k$ according to the definitions
\begin{subequations}
\begin{equation}\label{xLaxoperator}
\begin{array}{l}
X(x,t,k)=-ikJ+Q(x,t)\,, 
 \end{array} 
\end{equation}
\begin{equation}\label{tLaxoperator}
\begin{array}{lll}
T(x,t,k)&=&-2\gamma k\left[-ikJ+Q(x,t)\right]-2ikC^{(1)}+i\gamma \sigma
\left[Q^2(x,t)-Q_x(x,t)\right]+\\
\\
 & & -\sigma W+\sigma [C^{(1)},Q]+C^{(0)}\, .
 \end{array} 
\end{equation}
\end{subequations}
In these expressions the diagonal traceless constant matrix $J$ is
\begin{equation}\label{jmatrix}
J =\sigma-\frac{N^{(+)}-N^{(-)}}{N^{(+)}+N^{(-)}}\cdot\mathbf{1} =
\frac{2}{N^{(+)}+N^{(-)}}\cdot\left( \begin{array}{lc} N^{(-)} & \mathbf{0} \\
\mathbf{0} & -N^{(+)} \end{array} \right)\,. 
\end{equation}
In order to simplify the notation, from now on (and as already done in
the previous formula) we do not specify the dimension of the matrices
$\mathbf{0}$ and
$\mathbf{1}$, and we may even omit to write the matrix $\mathbf{1}$
altogether, as we trust the reader will not be confused by this omission.

%% Reductions
Let us now consider the condition 
\begin{equation}\label{matred}
Q^\dag(x,t)=S\,Q(x,t)\,S 
\end{equation}
on the solution $Q(x,t)$ of the matrix evolution equation (\ref{boomeron}) 
where the dagger stands for hermitian conjugation. The constant matrix
$S$ is block diagonal,
\begin{equation}\label{matsign}
S=\left(\begin{array}{cc}S^{(+)} & \mathbf{0} \\ \mathbf{0} & S^{(-)}\end{array}\right)\,; 
\end{equation}
its off--diagonal blocks are vanishing rectangular matrices while its
diagonal blocks $S^{(+)}$ and  $S^{(-)}$ are, respectively, 
$N^{(+)}\times N^{(+)}$ and $N^{(-)}\times N^{(-)}$ diagonal matrices
whose diagonal elements $s^{(\pm)}_n$, with no loss of generality, are
signs, namely
\begin{equation}\label{signs}
S^{(\pm)}=\mbox{diag}\,(s^{(\pm)}_1,\cdots,s^{(\pm)}_{N^{(\pm)}}\,)\,,\quad
{s^{(\pm)}_{n}}^2=1\,. 
\end{equation}
This of course implies the relations
$S^2=\mathbf{1}$,  $\,{S^{(+)}}^2=\mathbf{1}$, $\,{S^{(-)}}^2=\mathbf{1}$. 

The reduction equation (\ref{matred}) is well motivated by the fact that
it captures several interesting models of dispersive propagation of
multicomponent waves in weakly nonlinear media. In order to support this claim, we display some of such model
equations in the next subsection. To this aim it is convenient to rewrite the matrix
PDEs (\ref{boomeron}) in terms of the
 blocks $Q^{(+)}$, $Q^{(-)}$, 
$W^{(+)}$ and $W^{(-)}$, see (\ref{qwblock}). These read
\begin{subequations}
\label{mothersystem}
\begin{equation}\label{qmothersystem}
\begin{array}{lll} 
Q_{t}^{(\pm )} &=&C^{(0)(\pm )}\,Q^{(\pm )}-Q^{(\pm )}\,C^{(0)(\mp )}\pm 
\left[ C^{(1)(\pm )}\,Q_{x}^{(\pm )}-Q_{x}^{(\pm )}\,C^{(1)(\mp )}\right] \\
&\mp & \left[ W^{(\pm )}\,Q^{(\pm )}+Q^{(\pm )}\,W^{(\mp )}\right] 
\mp i\gamma \,~\left[ Q_{xx}^{(\pm )}-2\,Q^{(\pm )}\,Q^{(\mp )}\,Q^{(\pm )}\right], 
\end{array}
\end{equation}
\begin{equation}
W_{x}^{(\pm )}=\left[ C^{(1)(\pm )},\,Q^{(\pm )}\text{\thinspace }Q^{(\mp )}
\right] .  \label{wmothersystem}
\end{equation}
\end{subequations}
where $C^{(j)(+)},$ respectively $C^{(j)(-)},$ are the $N^{(+)}\times
N^{(+)},$ respectively $N^{(-)}\times N^{(-)},$ \textit{constant} square
matrix blocks of $C^{(j)}$, see (\ref{Cmatrices}).
The reduction condition (\ref{matred}) is accounted for by
introducing the dependent variable $U(x,t)$ through the definitions
\begin{equation}\label{reduction}
Q^{(-)}(x,t)=U(x,t)\,,\quad Q^{(+)}(x,t)=S^{(+)}U^{\dagger}(x,t)S^{(-)}\,.
\end{equation}
Since the constant $\gamma$ is real, these expressions of $Q^{(+)}$ and
$Q^{(-)}$ in terms of the single variable $U(x,t)$ are compatible with the
equations (\ref{mothersystem}) which then reduce to 
\begin{subequations}
\label{redboomeron}
\begin{equation}\label{Uboomeron}
\begin{array}{lll}
U_{t}& =&C^{(0)(-)}\,U-U\,C^{(0)(+)}-\left[ C^{(1)(-)}\,U_{x}-U_{x}
\,C^{(1)(+)}\right] +\\
&+& \left[ W^{(-)}\,U+U\,W^{(+)}\right] +i\gamma \,~\left[ U_{xx}-2\,U\,S^{(+)}\,U^{\dagger }\,S^{(-)}\,U\right] ,  
\end{array}
\end{equation}
\begin{equation}\label{Wminus}
W_{x}^{(+)}=\left[ C^{(1)(+)},\,S^{(+)}\,U^{\dagger }\,S^{(-)}\,U\right] ,
\end{equation}
\begin{equation}\label{plus}
W_{x}^{(-)}=\left[ C^{(1)(-)},U\,S^{(+)}\,U^{\dagger }\,S^{(-)}\right] .
\end{equation}
\end{subequations}
Here $U$ is an $N^{(-)}\times N^{(+)}$ rectangular matrix,
(see (\ref{reduction}) and (\ref{qwblock})), 
while the auxiliary variables $W^{(\pm )}(x,t)$ are square matrices, respectively $W^{(+)}$ is an $N^{(+)}\times N^{(+)}$ matrix and $W^{(-)}$ is an $N^{(-)}\times N^{(-)}$ matrix, and it is easily seen that they satisfy the ``hermitianity'' conditions
\begin{equation}\label{HermW}
W^{(+)}=-S^{(+)}\,W^{(+)\dagger}\,S^{(+)}\,,\quad 
W^{(-)}=-S^{(-)}\,W^{(-)\dagger }\,S^{(-)}\,,
\end{equation} 
that is
\begin{equation}\label{HermWa}
W^{\dagger}(x,t)=-S\,W(x,t)\,S\,.
\end{equation} 
Similarly, the constant matrices $C^{(j)(\pm )}$ satisfy
the following conditions: 
\begin{subequations}
\label{Cond1}
\begin{equation}\label{Cond1a}
C^{(j)(+)}=-\left( -\right) ^{j}\,S^{(+)}\,C^{(j)(+)\dagger
}\,S^{(+)}\,,\quad j=0,1,  
\end{equation}
\begin{equation}\label{Cond1b}
C^{(j)(-)}=-\left( -\right) ^{j}\,S^{(-)}\,C^{(j)(-)\dagger
}\,S^{(-)}\,,\quad j=0,1. 
\end{equation}
\end{subequations}
As for the ``sign'' matrices
$S^{(+)}$ and $S^{(-)}$, see (\ref{signs}), we note
that one could set, for instance, $s^{(+)}_1=1$ with no loss of
generality, but we prefer to keep the symmetrical, though redundant,
notation (\ref{signs}).  

% List of equations
\subsection{Multicomponent wave equations: cases of interest}\label{sec:list}
We observe that the general matrix nonlinear evolution  equation 
(\ref{redboomeron}) may well specialize itself to quite a large family of coupled wave equations
by playing with various choices of the integers $N^{(+)}$ and $N^{(-)}$, and of the constant matrix coefficients $C^{(j)(+)}$ and $C^{(j)(-)}$. This exercise is certainly worth doing since it happens that among this family of wave equations there are some which look interesting models in different applicative contexts. With this purpose in mind, and before proceeding further on our general setting, we briefly list here a few examples of model PDEs which we deem of interest in applications (for further details see also \cite{CD2004} and \cite{CD2006}). 

Let us begin with choosing
$N^{(+)}=1$ and $N^{(-)}=D$, $C^{(j)(\pm)}=0$ and let us set $S^{(+)}=1$ and
$S^{(-)}=\hat{S}=$diag$(s_1,\cdots ,s_D)$ with $s_n^2=1$. In
this case the dependent variable $U(x,t)$ is a (column) $D$--dimensional
vector
$\mathbf{u}(x,t)=(u^{(1)}(x,t),\cdots ,u^{(D)}(x,t))$ and the resulting
VNLS equation 
\begin{equation}\label{Dvector}
\mathbf{u}_t=i\gamma \left[{\mathbf{u}}_{xx}-2
\left(\sum_{n=1}^Ds_n|u^{(n)}|^2 \right)\mathbf{u}\right]=i\gamma ({\mathbf{u}}_{xx}-2
<\mathbf{u},\hat{S} \mathbf{u}>\mathbf{u})\,,
\end{equation}
is a simple generalization of the $D=2$ case (\ref{s1s2}).
Here and hereafter we adopt the bracket notation to indicate the (nonsymmetrical)
scalar product of two vectors, namely
\begin{equation}\label{scalar}
<\mathbf{u},\mathbf{v}>=\sum_{n=1}^D{u^{(n)}}^{\ast}v^{(n)}\,.
\end{equation}
This system of NLS equations can be further generalized by adding coupling terms which originate from
non vanishing coefficients $C^{(j)(\pm)}$. A simple instance of such generalization obtains for $N^{(+)}=1$ and $N^{(-)}=2$, together with $S^{(+)}=1$ and
$S^{(-)}=$diag$(s_1,s_2)$, where $s_{1,2}^2=1$. Here the choice of the matrix coefficients is
\begin{equation}\label{boomNLScoeff}
C^{(0)(+)}=C^{(1)(+)}=0,\quad
C^{(0)(-)}=\left(\begin{array}{cc} 0 & s_1 a\\  -s_2 a & 0\end{array}\right)\,, \quad C^{(1)(-)}=\left(\begin{array}{cc} -b & 0 \\ 0 & b\end{array}\right)\,\,,
\end{equation}
while the dependent variables are the two components  
$(u^{(1)}(x,t)\, ,\,u^{(2)}(x,t))$ of the 2--dimensional (column) vector $U$
 and the function $w(x,t)$ is defined through the off--diagonal matrix $W^{(-)}$ 
 \begin{equation}\label{boomNLSw} 
 W^{(-)}=\left(\begin{array}{cc} 0 & -s_1 w\\  s_2 w^{\ast }& 0\end{array}\right)
\end{equation}
while $W^{(+)}=0$. With these specifications the VNLS equations read
\setlength\arraycolsep{2pt}
\begin{equation}\label{boomNLS}
\begin{array}{l}
u^{(1)}_{t} =s_{1}\, a u^{(2)}+b u^{(1)}_{x}-s_{1}\,w u^{(2)} +i \gamma \left[ u^{(1)}_{xx}-2\left( s_{1}\,\left\vert u^{(1)}\right\vert
^{2}+s_{2}\,\left\vert u^{(2)}\right\vert ^{2}\right) u^{(1)}\right] ,\\
u^{(2)}_{t} =-s_{2}\,a^{\ast } u^{(1)}-b u^{(2)}_{x}+s_{2}\,w^{\ast } u^{(1)} +i \gamma\left[ u^{(2)}_{xx}-2\left( s_{1}\,\left\vert u^{(1)}\right\vert
^{2}+s_{2}\,\left\vert u^{(2)}\right\vert ^{2}\right) u^{(2)}\right] ,\\
\,w_{x}=2\,b\,s_{1}\,s_{2}\,u^{(1)}\,u^{(2)\ast }\,,
\end{array}
\end{equation}
where $a$ is an arbitrary  complex coefficient and $b$ is an arbitrary real coefficient. We note that this system results from a mixing of Schr\"{o}dinger--type dispersion and quadratic nonlinearity as it occurs in the 3-wave resonant interaction. 
This extension is simple and it might look trivial; however, because of the non commutative character of the matrix coefficients $C^{(0)(-)}$ and $C^{(1)(-)}$, which entails the introduction of the new auxiliary dependent variable $w(x,t)$, it introduces the interesting phenomenology of boomerons, i.e. solitons which have different asymptotic velocities at $t=+\infty$ and $t=-\infty$. These special soliton solutions were first introduced long time ago in \cite{CD1976} (see also 
\cite{D1978}) in connection with the hierarchy of the matrix Korteweg--de Vries equation, but only recently boomerons appeared again in geometry \cite{DRS2002} and optics \cite{DCBW2006}, \cite{mantsyzov}. 

Let us consider next $N^{(+)}=2$, $N^{(-)}=D$, $C^{(j)(\pm)}=0$, $S^{(+)}=\,$diag$(s_1,s_2)$ and $S^{(-)}=\hat{S}=\,$diag$(s_1,\cdots ,s_D)$. In this case the matrix $U$ has
two columns, namely two $D$--dimensional vectors, $\mathbf{u}$ and
$\mathbf{v}$. They satisfy the following coupled VNLS equations
\begin{equation}\label{2Dvector} 
\begin{array}{l}
\mathbf{u}_t=i\gamma
({\mathbf{u}}_{xx}-2s^{(+)}_1<\mathbf{u},S^{(-)}
\mathbf{u}>\mathbf{u}-2s^{(+)}_2<\mathbf{v},S^{(-)}
\mathbf{u}>\mathbf{v})~,\\  \mathbf{v}_t=i\gamma
({\mathbf{v}}_{xx}-2s^{(+)}_1<\mathbf{u},S^{(-)}
\mathbf{v}>\mathbf{u}-2s^{(+)}_2<\mathbf{v},S^{(-)}
\mathbf{v}>\mathbf{v})~.
\end{array}
\end{equation}
This system may of course be rewritten as one equation for a
$2D$--dimensional vector. However, this vector equation would be different
from the equation which obtains in the case $N^{(+)}=1$, $N^{(-)}=2D$. Of course, similarly to the previous case, also this system can be generalized so as to feature boomeronic--type effects.  

Consider now the subcase $D=2$, namely $N^{(+)}=N^{(-)}=2$. 
The matrix $U$ is $2\times2$ square and therefore it can be given a
different representation by using Pauli matrices:
\begin{equation}\label{pauli}
U=iu^{(4)}\mathbf{1}+u^{(1)}\sigma_1+u^{(2)}\sigma_2+u^{(3)}\sigma_3\,;
\end{equation} 
with this parametrization of $U$ the four dependent scalar variables are
the components $u^{(j)}, j=1,\cdots,4,$ of the 4--dimensional vector
$\vec{u}=(u^{(1)}, u^{(2)}, u^{(3)}, u^{(4)})$. However, the resulting
system of evolution equations for these fields takes a vector covariant
form by specializing the ``sign" matrices to
$S^{(+)}=\mathbf{1}, S^{(-)}=s\mathbf{1}$ with $s^2=1$. In this case, the
evolution equation for the 4--dimensional complex vector $\vec{u}$ has been
first introduced in
\cite{CD2004}, and reads, in self--evident notation,
\begin{equation}\label{paulinls}
\vec{u}_t=i\gamma\{\vec{u}_{xx}-2s[2
(\vec{u}^{\ast}\cdot\vec{u})\vec{u}-(\vec{u}\cdot\vec{u})
\vec{u}^\ast ]\}\,.
\end{equation} 
Let us note that, via the transformation $u^{(j)}\rightarrow
u^{(j)}\exp(i\phi_j)$ with $\phi_j$ four arbitrary real constants,
this system (\ref{paulinls}) of four coupled NLS equations takes the
form
\begin{equation} \label{generalpauli}
u^{(k)}_t=i\gamma\left\{u^{(k)}_{xx}-2s\left[2
\left(\sum_{j=1}^4|u^{(j)}|^2 \right)u^{(k)}-\left(\sum_{j=1}^4\exp(i\delta_{jk}){u^{(j)}}^2 \right){u^{(k)}}^\ast \right] \right\}\,,
\end{equation} 
where the real constants  
$\delta_{jk}={\delta_{jk}}^{\ast}=-\delta_{kj}$, with
$\delta_{jk}=2(\phi_j-\phi_k)$, are three
arbitrary phases. We also note that further reductions of this system
are obtained by merely letting one, two or three components of the 4--vector
$\vec{u}$ vanish. In fact, by setting for instance
$u^{(4)}=0$, the system (\ref{generalpauli}) becomes a 3--dimensional
VNLS equation with two arbitrary phases, say $\delta_{21}$ and
$\delta_{32}$. Setting also $u^{(3)}=0$ leads then to the 2--dimensional case, which is 
\begin{equation}\label{delta}
\begin{array}{ll}
u^{(1)}_t=i\gamma\{u^{(1)}_{xx}-2s[(|u^{(1)}|^2+2|u^{(2)}|^2)u^{(1)}+
\exp(i\delta)u^{(2)2}u^{(1)\ast}]\}\,,& \\ 
u^{(2)}_t=i\gamma\{u^{(2)}_{xx}-2s[(2|u^{(1)}|^2+|u^{(2)}|^2)u^{(2)}+
\exp(-i\delta)u^{(1)2}u^{(2)\ast}]\}\,,&
\end{array}
\end{equation}
 with only one arbitrary constant phase, $\delta_{21}=\delta$. This system has been first introduced in \cite{FK83} with $\delta=0$,
while the two special cases with $\delta=0$ and $\delta=\pi$
had been identified as the only integrable ones in the class of
equations having the form written above, but with the two factors
$\exp(i\delta)$ and $\exp(-i\delta)$ replaced by two apriori
arbitrary real constants \cite{Khukhunashvili89}. The case in which three components of $\vec{u}$ are vanishing yields the scalar NLS
equation (\ref{nls}). Larger systems of coupled nonlinear NLS equations
are similarly obtained by starting with a bigger square matrix $U(x,t)$.
For instance 9--dimensional VNLS equations result from expanding the
$3\times3$ matrix $U$ in the basis obtained by adding the unit matrix to
the SU(3) group generators. Other ways of coupling NLS equations may be
deviced by different parametrizations of the rectangular matrix $U$.

Finally, let us note that the well known 3--wave resonant interaction equation (in both its ``explosive" and ``non--explosive" versions) is a special case of our family of reduced equations as it merely coincides with the $a= \gamma=0$ case of the system (\ref{boomNLS})
\begin{equation}\label{3wri}
\begin{array}{l}
u^{(1)}_{t} -b\,u^{(1)}_{x}= -s_{1}\,\,w\,u^{(2)}  ,\\
u^{(2)}_{t} +b\,u^{(2)}_{x}= s_{2}\,\,w^{\ast }\,u^{(1)} ,\\
\,w_{x}=2\,b\,s_{1}\,s_{2}\,\,u^{(1)}\,u^{(2)\ast }\,.
\end{array}
\end{equation}
Moreover, having its physical applications in mind, one should also redefine for this equation the two independent variables $x$ and $t$ rather as ``time" and, respectively,  ``space", with the implication that the three characteristic group velocities are  $1/b$, $-1/b$ and $0$. Indeed, this observation has played a basic role in the discovery \cite{CD3wave} of soliton solutions of the 3WRI equation such as boomerons, trappons, simultons \cite{DCBW2006} and pair creation.

%% DDT
\section{The Darboux-Dressing Transformation}\label{sec:DDT}     
Let us now turn our attention to the method of construction of special
solutions of the general system (\ref{redboomeron}). We first
note that the reduction conditions (\ref{matred}), (\ref{HermWa}), (\ref{Cond1}), together with the expressions (\ref{xLaxoperator}) and (\ref{tLaxoperator}) of the matrices
$X(x,t,k)$ and
$T(x,t,k)$ in the two linear equations (Lax pair)
(\ref{lax}) entail the following relations:
\begin{equation}\label{Laxred}
X^{\dagger}(k^{\ast})\Sigma+\Sigma
X(k)=0\,,\quad T^{\dagger}(k^{\ast})\Sigma+\Sigma T(k)=0\,,
\end{equation}
with 
\begin{equation}\label{Sigma}
\Sigma=\sigma S=\left(\begin{array}{cc}S^{(+)} &
\mathbf{0}  \\ \mathbf{0}  &
-S^{(-)}\end{array}\right)\,,\quad\Sigma^2=\mathbf{1}\,.
\end{equation}
As for the notation used in (\ref{Laxred}), we have omitted writing the
dependence on the variables $x$ and $t$ and we maintain this
omission in the following whenever it will cause no confusion. The
property (\ref{Laxred}) allows one to express the reduction condition induced
by (\ref{matred}), (\ref{HermWa}), (\ref{Cond1})  on the solution $\psi(k)$ of the two linear equations
(\ref{lax}); this condition is given by the following equation:
\begin{equation}\label{psired}
\psi^{\dagger}(x,t,k^\ast)\Sigma \psi(x,t,k)=A(k,k^\ast) \,,
\end{equation}
where the matrix $A(k,k^\ast)$ is constant, namely $x$-- and
$t$--independent. Therefore, it is plain that the value of
$A(k,k^\ast)$  depends only on the 
arbitrary value $\psi(x_0,t_0,k)$ that the solution
$\psi$ takes at a  given point $(x_0,t_0)$ of the $(x,t)$ plane.

Consider now a second pair of matrices $Q^{(0)}(x,t)$ and $W^{(0)}(x,t)$, and assume that they have the same block structure of $Q$ and $W$, see
(\ref{qwblock}), and satisfy the same reduction conditions (\ref{matred}), (\ref{HermWa}).
Let $\psi^{(0)}(x,t,k)$ be a corresponding
nonsingular (i.e. with nonvanishing determinant) matrix solution of (\ref{lax})
\begin{equation}\label{0Laxpair}
\psi^{(0)}_x=X^{(0)}\psi^{(0)}\,,
\quad \psi^{(0)}_t=T^{(0)}\psi^{(0)}
\end{equation}
with $X^{(0)}(x,t,k)$ and
$T ^{(0)}(x,t,k)$ having the expressions (\ref{xLaxoperator}) and
(\ref{tLaxoperator}) with $Q$ and $W$ replaced by
$Q^{(0)}$ and $W^{(0)}$. Assume also that the initial condition $\psi^{(0)}(x_0,t_0,k)$
is so chosen that the constant matrix $A^{(0)}(k,k^\ast)$, where of
course (see (\ref{psired}))
\begin{equation}\label{0psired}
 A^{(0)}(k,k^\ast)={\psi^{(0)}}^{\dagger}(x,t,k^\ast)\Sigma
\psi^{(0)}(x,t,k)\,,
\end{equation}
coincides with $A(k,k^\ast)$, i.e. $A^{(0)}(k,k^\ast)=A(k,k^\ast)$.  
 Since both compatibility conditions,
$\psi^{(0)}_{xt}=\psi^{(0)}_{tx}$ and $\psi_{xt}=\psi_{tx}$, are
satisfied, $Q^{(0)}(x,t)$, $W^{(0)}(x,t)$ and $Q(x,t)$, $W(x,t)$ are two different solutions of the same matrix evolution equations (\ref{boomeron}), and therefore it
follows that the matrix
\begin{equation}\label{Dmatrix}
D(x,t,k)=\psi(x,t,k)(\psi^{(0)}(x,t,k))^{-1}
\end{equation}  
satisfies the differential equations
\begin{equation}\label{Ddiff}
D_x=XD-DX^{(0)}\,,\quad D_t=TD-DT^{(0)}\,, 
\end{equation}
 together with the
algebraic (reduction) equation
\begin{equation}\label{Dalgeb}
D^\dagger(k^\ast)\Sigma D(k)=\Sigma\,.
\end{equation}
The proof of these propositions is straight.

The definition (\ref{Dmatrix}) can be viewed as a transformation of
$\psi^{(0)}$ into $\psi$
\begin{equation}\label{DDT}
\psi(x,t,k)=D(x,t,k)\psi^{(0)}(x,t,k)\,, 
\end{equation}
which consequently yields a transformation of $Q^{(0)}$ and $W^{(0)}$ into, respectively,  $Q$ and 
$W$.
Therefore, the dressing approach requires in the first place an explicit 
knowledge of $Q^{(0)}(x,t)$, $W^{(0)}(x,t)$ and $\psi^{(0)}(x,t,k)$. The next
step is the construction of the transformation matrix $D(x,t,k)$ via
the integration of the ODEs (\ref{Ddiff}). This task is however not
straight since the coefficients $X$ and $T$ of these differential
equations depend on the unknown matrices $Q$ and $W$ (see (\ref{xLaxoperator})
and (\ref{tLaxoperator})). The way of solving this problem goes through
the \emph{a priori} assignment of the dependence of the
transformation matrix $D(x,t,k)$ on the spectral variable $k$. 

In the following we will investigate the
set of $k$--dependent matrices $D(k)$ which i) have a
\emph{rational} dependence on the complex variable $k$ and ii) have
\emph{nonvanishing} $k\rightarrow \infty$ limit. Moreover, if we 
consider a
rational dependence on $k$ which can be factorized as product of simple-pole
terms, we need to deal only with matrices $D(k)$ which take the following
one-pole expression
\begin{equation}\label{DDTmat}
D(x,t,k)=\mathbf{1}+\frac{R(x,t)}{k-\alpha}\,,
\end{equation}  
where the matrix $R(x,t)$ is the residue at the pole $k=\alpha$ and the value of $D(k)$ in the $k\rightarrow \infty$ limit is taken to be the identity for the sake of simplicity. The transformation (\ref{DDT}) characterized by the matrix (\ref{DDTmat}) has 
received considerable attention in the literature \cite{DDT1}--\cite{DDT4} ( see also \cite{DL}).
 We refer to it as   
Darboux--Dressing Transformation (DDT), and  
its existence in our setting is proved below by construction.

The way to obtain an explicit expression of the residue matrix $R(x,t)$
depends on whether the pole $\alpha$ is off the real axis,
$\alpha \neq \alpha^\ast$, or on the real axis, $\alpha = \alpha^\ast$. Therefore we treat these two cases separately.

%% alpha complex
\subsection{The Darboux-Dressing Transformation: complex pole}\label{sec:DDTcompl} 
 Let us begin with considering the case in
which $\alpha$ is not real, $\alpha\neq\alpha^\ast$, while the other case, $\alpha$ real, is discussed in the next subsection \ref{sec:DDTreal}.  The starting
point is the requirement that the matrix $D(x,t,k)$, see
(\ref{DDTmat}), satisfies the algebraic condition (\ref{Dalgeb}) and
the differential equations (\ref{Ddiff}). The algebraic condition
entails the two (equivalent) equations
\begin{equation}\label{algeq}
\Sigma R+\frac{R^\dagger \Sigma R}{\alpha-\alpha^\ast}=0\,,
\quad R^\dagger \Sigma -\frac{R^\dagger \Sigma
R}{\alpha-\alpha^\ast}=0\,,
\end{equation}  
whose solution is 
\begin{equation}\label{residue}
R(x,t)=(\alpha-\alpha^\ast)P(x,t)\,,
\end{equation}
where the matrix $P(x,t)$ is a projector with the ``hermitianity" condition
\begin{equation}\label{projector}
P^2=P\,,\quad P^\dagger=\Sigma P \Sigma\,. 
\end{equation}
As for the differential equations (\ref{Ddiff}), replacing $D(x,t,k)$
with its expression (\ref{DDTmat}) (and (\ref{residue})) yields the algebraic relations
\begin{subequations}\label{Backlund}
\begin{equation}\label{qBacklund}
Q=Q^{(0)}+i(\alpha-\alpha^\ast)[\sigma,P]\,,
\end{equation}
\begin{equation}\label{wBacklund}
W=W^{(0)}-i(\alpha-\alpha^\ast)[C^{(1)},\{\sigma,P\}]\,,
\end{equation}
\end{subequations}
which give the ``dressed" matrices $Q$ and $W$ in terms of the ``bare" matrices
$Q^{(0)}$, $W^{(0)}$ and the projector $P$, together with the two differential equations
\begin{equation}\label{diffeq}
P_x=X(\alpha)P-PX^{(0)}(\alpha)\,,\quad P_t=T(\alpha)P-PT^{(0)}(\alpha)\,,
\end{equation}
whose integration goes as follows. Consider first the differential
equation (\ref{diffeq}) with respect to the variable $x$ and replace $Q$
with its expression (\ref{qBacklund}). The resulting equation is then the
following nonlinear equation
\begin{equation}\label{Pxdiff}
P_x=X^{(0)}(\alpha)P-PX^{(0)}(\alpha)+i(\alpha-\alpha^\ast)(JP-PJP)
\,.
\end{equation}
Let $z$ be an eigenvector of P, and differentiate with respect to $x$
the eigenvalue equation 
\begin{equation}\label{eigenvector}
Pz=z\,.
\end{equation}
By replacing then $P_x$ with the right--hand side of (\ref{Pxdiff}), one
arrives to the equation
\begin{equation}
(\mathbf{1}-P)[z_x-X^{(0)}(\alpha^\ast)z]=0\,,\label{zxdiff}
\end{equation}
which implies that the vector $z_x-X^{(0)}(\alpha^\ast)z$ is in the
subspace on which $P$ projects. At this point we may well assume that
this subspace is one--dimensional. Indeed, it is easy to prove that, if
$P$ projects on a subspace of higher dimension $n>1$, then the
corresponding matrix $D$ is a product of as many matrices $D^{(j)}$ of
the form (\ref{DDTmat}) with (\ref{residue}),
\begin{equation}\label{jDDT}
D^{(j)}(x,t,k)=\mathbf{1}+\frac{(\alpha-\alpha^\ast)
P^{(j)}(x,t)}{k-\alpha}\,,\quad j=1,\cdots,n\,,
\end{equation}
as the dimension $n$ of this subspace,
all of course with the same pole in $\alpha$, and all with 
$P^{(j)}$ projecting on a one--dimensional subspace. Therefore, with no
loss of generality, we let $P$ in the DDT matrix
\begin{equation}\label{elementaryDDT}
D(x,t,k)=\mathbf{1}+\frac{(\alpha-\alpha^\ast)
P(x,t)}{k-\alpha}\,,
\end{equation}
project on the one--dimensional
subspace of the vector $z$,
with the implication, see (\ref{zxdiff}), that the vector
$z_x-X^{(0)}(\alpha^\ast)z$ is proportional to $z$. On the other hand,
since the vector $z$ is defined here modulo a scalar factor function, we
may choose this factor in such a way that $z$ satisfies the differential
equation
\begin{equation}\label{zxeq}
z_x=X^{(0)}(\alpha^\ast)\,z\,\,. 
\end{equation}
The differential equation (\ref{diffeq}) with respect to the variable
$t$ can be treated in a similar way. The substitution of $Q$ and $W$ with their 
expressions (\ref{Backlund}) yields the nonlinear equation
\begin{equation} \label{Ptdiff}
\begin{array}{lll}
P_t &=& T^{(0)}(\alpha)P-PT^{(0)}(\alpha)+2\gamma
(\alpha-\alpha^\ast)(Q^{(0)}P-PQ^{(0)}P)+\\
&-& 2i\gamma(\alpha^2-\alpha^\ast{}^2)(JP-PJP)+2i(\alpha-\alpha^\ast)[C^{(1)},P]P\,.
\end{array}
\end{equation}
By differentiating now the eigenvalue equation (\ref{eigenvector}) with
respect to $t$ and using both the equations (\ref{eigenvector}) and
(\ref{Ptdiff}),  one ends up with the equation
\begin{equation}\label{ztdiff}
(\mathbf{1}-P)[z_t-T^{(0)}(\alpha^\ast)z]=0\,,
\end{equation}
which, by the same arguments as above, implies that the vector $z(x,t)$
satisfies the differential equation
\begin{equation}\label{zteq}
z_t=T^{(0)}(\alpha^\ast)\,z\,.
\end{equation}
Once the two equations (\ref{zxeq}) and (\ref{zteq}) are solved, the DDT
transformation matrix $D(x,t,k)$ is finally given by
(\ref{elementaryDDT}) with
\begin{equation}\label{diadic}
P(x,t)=\frac{z(x,t)z^\dagger(x,t)\Sigma}{<z(x,t),\Sigma
z(x,t)>}\,,
\end{equation}
this expression being implied by the algebraic conditions
(\ref{projector}).
At this point we conclude that the method of construction of a novel 
solution $Q(x,t)$, $W(x,t)$ of the evolution equations (\ref{boomeron}), starting from
the knowledge of given (\emph{seed}) solution $Q^{(0)}(x,t)$, $W^{(0)}(x,t)$ is explicitly given
by (\ref{Backlund}) with (\ref{diadic}) where the
vector $z(x,t)$ is 
\begin{equation}\label{zsolution}
z(x,t)=\psi^{(0)}(x,t,\alpha^\ast)\,z_0 \,. 
\end{equation}
Here $\psi^{(0)}(x,t,\alpha^\ast)$ is the solution
$\psi^{(0)}(x,t,k)$  of the differential equations
(\ref{0Laxpair}) (Lax pair  corresponding to $Q^{(0)}(x,t)$, $W^{(0)}(x,t)$), for $k=\alpha^\ast$,
and it is assumed to be known, while
$z_0$ is an arbitrary constant $(N^{(+)} \,+\,N^{(-)})$--dimensional vector.

The construction of the novel solution $Q(x,t)$, $W(x,t)$  provides, via the 
formulas (\ref{qwblock}) and the reduction (\ref{reduction}),  the
construction of the novel solution $U(x,t)$, $W^{(\pm)}(x,t)$ of the matrix evolution equation
(\ref{redboomeron}) which is our main concern here. The relevant
expressions are obtained by first writing the block structure of the
$(N^{(+)}+N^{(-)})\times (N^{(+)}+N^{(-)})$ projector matrix $P(x,t)$
\begin{equation}\label{Pblock}
P=\left(\begin{array}{cc} B^{(+)} & -S^{(+)}BS^{(-)} \\ B & B^{(-)}\end{array}\right)\,,
\end{equation}   
which therefore entails, in self--evident notation and by using
(\ref{Backlund}), the relations
\begin{subequations}\label{uwpmBacklund}
\begin{equation}\label{uBacklund}
U=U^{(0)}-2i(\alpha-\alpha^\ast)B\,,
\end{equation}
\begin{equation}\label{wpBacklund}
W^{(+)}=W^{(0)(+)}-2i(\alpha-\alpha^\ast)[C^{(1)(+)},B^{(+)} ]\,,
\end{equation}
\begin{equation}\label{wmBacklund}
W^{(-)}=W^{(0)(-)}+2i(\alpha-\alpha^\ast)[C^{(1)(-)},B^{(-)}]\,.
\end{equation}
\end{subequations}
In view of its use in future computations, we give to this formula a more
explicit expression by using the form (\ref{diadic}) of the projector $P$.
To this aim it is convenient to split the vector $z$ in two block 
column vectors, namely
\begin{equation}\label{zblock}
z=\left(\begin{array}{c}z^{(+)}  \\
z^{(-)} \end{array}\right)\,,
\end{equation}
where the vectors $z^{(+)}$ and $z^{(-)}$ have dimension $N^{(+)}$ and, respectively, 
$N^{(-)}$. Then, by inserting this block form of $z$ in the diadic
expression of $P$, (\ref{diadic}) (and recalling (\ref{Sigma})), we arrive
to the final relations
\begin{subequations}\label{Backlundcompl}
\begin{equation}
U=U^{(0)}-2i(\alpha-\alpha^\ast)\frac{z^{(-)}z^{(+)}
{}^\dagger S^{(+)}}{<z^{(+)},S^{(+)}z^{(+)}>-<z^{(-)},S^{(-)}z^{(-)}>}\,,
\label{uBacklundcompl}
\end{equation} 
\begin{equation}
W^{(+)}=W^{(0)(+)}-2i(\alpha-\alpha^\ast)\frac{(C^{(1)(+)}\, z^{(+)}z^{(+)}
{}^\dagger S^{(+)}-z^{(+)}z^{(+)}
{}^\dagger S^{(+)}\,C^{(1)(+)})}{<z^{(+)},S^{(+)}z^{(+)}>-<z^{(-)},S^{(-)}z^{(-)}>}\,,
\label{wpBacklundcompl}
\end{equation} 
\begin{equation}
W^{(-)}=W^{(0)(-)}+2i(\alpha-\alpha^\ast)\frac{(C^{(1)(-)}\, z^{(-)}z^{(-)}
{}^\dagger S^{(-)}-z^{(-)}z^{(-)}
{}^\dagger S^{(-)}\, C^{(1)(-)})}{<z^{(+)},S^{(+)}z^{(+)}>-<z^{(-)},S^{(-)}z^{(-)}>}\,.
\label{wmBacklundcompl}
\end{equation} 
\end{subequations}

%% alpha real
\subsection{The Darboux-Dressing Transformation: real pole}\label{sec:DDTreal}
Let us now investigate the alternative case in which the pole of the
Darboux--Dressing matrix $D(x,t,k)$, see (\ref{Dmatrix}), is real,
$\alpha=\alpha^\ast$. The way to treat this case is the same as that one
we followed in the previous case, but the resulting equations to be
solved are indeed different. Thus we first ask that $D(x,t,k)$, as given
by the general expression (\ref{DDTmat}), satisfies both the algebraic
condition (\ref{Dalgeb}) and the differential equations (\ref{Ddiff}).
The algebraic condition implies two equations for the residue matrix
$R$:
\begin{equation}\label{realalgeq}
\Sigma R+R^\dagger \Sigma=0\,,\quad R^\dagger \Sigma R=0\,.
\end{equation}
These entail the following form of $R$
\begin{equation}\label{residuereal}
R(x,t)=i\rho(x,t)\, \hat{P}(x,t)\Sigma \,,
\end{equation}
together with the conditions that the scalar function $\rho(x,t)$ is real,
the projector matrix $\hat{P}$ is hermitian,
\begin{equation}\label{reality}
\rho=\rho^\ast\,,\quad\hat{P}^2=\hat{P}\,,\quad\hat{P}=
\hat{P}^\dagger\,,
\end{equation}
and it satisfies the equation
\begin{equation}\label{Porthog}
\hat{P}\Sigma\hat{P}=0\,.
\end{equation}
Therefore, in the present case, the transformation matrix (\ref{DDTmat})
reads
\begin{equation}\label{realDDT}
D(x,t)=\mathbf{1}+i\rho(x,t)\frac{\hat{P}(x,t)
\Sigma}{k-\alpha}\,.
\end{equation} 
By the same arguments we have used in the previous case, one
can show tha $\hat{P}$ may be assumed, with no loss of generality, to
project on a one--dimensional subspace, namely
\begin{equation}\label{diadicreal}
\hat{P}=\frac{\hat{z}
\hat{z}^\dagger}{<\hat{z},\hat{z}>}\,,
\end{equation} 
where the vector $\hat{z}(x,t)$, because of the equation
(\ref{Porthog}), is constrained by the orthogonality condition
\begin{equation}\label{orthogonality}
<\hat{z},\Sigma \hat{z}>=0\,.
\end{equation}
Consider now the differential equations (\ref{Ddiff}) and insert in
them the expression (\ref{realDDT}). Since $k$ is of course an
arbitrary complex variable, we obtain the relations
\begin{subequations}\label{realBacklund}
\begin{equation}\label{qrealBacklund}
 Q=Q^{(0)}+\rho[\hat{P},\sigma]\Sigma\,,
\end{equation}  
\begin{equation}\label{wrealBacklund}
 W=W^{(0)}+\rho[C^{(1)},\{\sigma,\hat{P}\}\Sigma]\,,
\end{equation} 
\end{subequations} 
which give the novel solution $Q(x,t)$, $W(x,t)$ of (\ref{boomeron}) in terms of the
supposedly known solution $Q^{(0)}(x,t)$, $W^{(0)}(x,t)$, the function $\rho(x,t)$ and the projector $\hat{P}(x,t)$. There also follows the relation
\begin{equation}\label{instru}
T(\alpha)-T^{(0)}(\alpha)=4\gamma \alpha
\rho[\sigma,\hat{P}]\Sigma+2i\gamma\rho[Q^{(0)},\hat{P} \Sigma] +2i\gamma
\rho^2\frac{g}{f}\hat{P}\Sigma-2\rho\sigma[C^{(1)},\sigma\hat{P}\Sigma]\,,
\end{equation}
which is not an independent relation but it is instrumental in deriving the
formula (\ref{realPtdiff}) displayed below. In addition, one obtains two following
differential equations, one with respect to $x$:
\begin{equation}\label{realPxdiff}
(\rho\hat{P})_x=\rho\left(X^{(0)}(\alpha)\hat{P}+\hat{P}
X^{(0)}{}^\dagger(\alpha)+\rho \frac{g}{f}\hat{P}\right)\,,
\end{equation}
and one with respect to $t$:
\begin{equation}\label{realPtdiff}
(\rho\hat{P})_t=\rho\left(T^{(0)}(\alpha)\hat{P}+\hat{P}
T^{(0)}{}^\dagger(\alpha)-4\alpha\gamma \rho 
\frac{g}{f}\hat{P}-2i\gamma\rho\frac{h}{f}\hat{P}+2\rho\frac{m}{f}\hat{P}\right)\,.
\end{equation}
In these  last three equations we have conveniently introduced the functions
$f(x,t)$, $g(x,t)$, $h(x,t)$ and $m(x,t)$ according to the definitions
\begin{equation}\label{functions}
f=<\hat{z},\hat{z}>\,,\quad g=<\hat{z},S\hat{z}>\,,\quad h=<\hat{z},\Sigma
Q^{(0)}\hat{z}>\,,\quad m=<\hat{z},\Sigma C^{(1)} \hat{z}>\,.
\end{equation} 
We have obtained the right--hand side of the differential equation
(\ref{realPtdiff}) by using the relation (\ref{instru}). At this point we
differentiate the eigenvalue equation (see (\ref{diadicreal}))
\begin{equation}\label{realeigenvector}
\hat{P} \hat{z}=\hat{z}\,.
\end{equation}
When this is done with respect to $x$, one obtains the equation
\begin{equation}
\hat{P}[\hat{z}_x+X^{(0)}(\alpha){}^\dagger\hat{z}]=\hat{z}_x-
X^{(0)}(\alpha)\hat{z}+\left(\frac{\rho_x}{\rho}-\rho 
\frac{g}{f}\right)\hat{z}\,,
\label{primazxdiff}
\end{equation}
which implies that its right--hand side is proportional to the vector $\hat{z}$,
\begin{equation}\label{propo}
\hat{z}_x-
X^{(0)}(\alpha)\hat{z}+\left(\frac{\rho_x}{\rho}-\rho \frac{g}{f}\right)\hat{z}=\mu
\hat{z}\,.
\end{equation}
On the other hand, the vector $\hat{z}$, which has been introduced through
the diadic expression (\ref{diadicreal}), is defined only modulo a factor
scalar function, and therefore, by taking advantage of this freedom, one
can ask that
\begin{equation}\label{mufix}
\mu=\frac{\rho_x}{\rho}-\rho \frac{g}{f}\,, 
\end{equation}
with the implication that the vector $\hat{z}$ satisfies the differential
equation
\begin{equation}\label{realzxdiff}
\hat{z}_x=X^{(0)}(\alpha)\hat{z}\,.
\end{equation}
We also observe that the function $\mu$, via its own definition
\begin{equation}\label{mudef}
\hat{P}[\hat{z}_x+X^{(0)}(\alpha){}^\dagger\hat{z}]=\mu \hat{z}\,\,,
\end{equation}
takes the following expression 
\begin{equation}\label{muexp}
\mu =\frac1f <\hat{z}\,,\,\hat{z}_x+X^{(0)}(\alpha){}^\dagger\hat{z}>=\frac1f(<\hat{z}\,,\,\hat{z}_x>\,+\,<\hat{z}_x\,,\,\hat{z}>)=\frac{f_x}{f}\,\,,
\end{equation}
where we have taken into account the equation (\ref{realzxdiff}) and the definition of $f$ (\ref{functions}).
It is now readily found that combining this equation with the relation (\ref{mufix}) entails that  the function $\rho(x,t)$ satisfies the following differential equation 
\begin{equation}\label{roxdiff}
\left(\frac{f}{\rho}\right)_x=-g\,.
\end{equation}
Let us now differentiate the eigenvalue equation (\ref{realeigenvector})
with respect to $t$. Following the same strategy as before, one concludes that the vector $\hat{z}$ satisfies the differential equation
\begin{equation}\label{realztdiff}
\hat{z}_t=T^{(0)}(\alpha)\hat{z}\,,
\end{equation} 
while the following relations hold true
\begin{equation}\label{nudef} 
\hat{P}[\hat{z}_t+T^{(0)}(\alpha){}^\dagger\hat{z}]=\nu \hat{z}\,\,,
\end{equation}
\begin{equation} \label{nufix}
\nu=\frac{\rho_t}{\rho}+2\frac{\rho}{f} \left(
2\alpha\gamma g+i\gamma h-m\right)\,\,.
\end{equation}
Again, the equation (\ref{nudef}) implies the expression
\begin{equation}\label{nuexp}
\nu =\frac1f <\hat{z}\,,\,\hat{z}_t+T^{(0)}(\alpha){}^\dagger\hat{z}>=\frac1f(<\hat{z}\,,\,\hat{z}_t>\,+\,<\hat{z}_t\,,\,\hat{z}>)=\frac{f_t}{f}\,\,,
\end{equation}
which, together with (\ref{nufix}), yields the differential equation
\begin{equation}\label{rotdiff}
\left(\frac{f}{\rho}\right)_t=2(2\alpha\gamma g+i\gamma h-m)
\end{equation}
for the function $\rho(x,t)$ with respect to the variable $t$.
Since the vector $\hat{z}(x,t)$ satisfies the two (compatible)
differential equations (\ref{realzxdiff}) and (\ref{realztdiff}), its
general expression is 
\begin{equation}\label{realzsolution}
\hat{z}(x,t)=\psi^{(0)}(x,t,\alpha)~\hat{z}_0\,, 
\end{equation}
where $\hat{z}_0$ is an arbitrary constant $(N^{(+)} \,+\,N^{(-)})$--dimensional vector.
The two differential equations (\ref{roxdiff}) and
(\ref{rotdiff}), which are also compatible with each other (the proof is straightforward and it is not reported here), can be easily integrated since the functions $f,g,h$ and $m$ are known (see (\ref{functions})). The  expression of their general solution then reads
\begin{equation}\label{roexpl}
\rho(x,t)=\frac{f(x,t)}{\left\{ \frac{f(x_0,t_0)}{\rho(x_0,t_0)}+2\int_{t_0}^tdt' 
\left[2\alpha\gamma g(x_0,t')+ i\gamma h(x_0,t')-m(x_0,t')\right]-\int_{x_0}^xdx' g(x',t)
\right\}}\,,
\end{equation}
where $x_0\,,\,t_0$ and $\rho(x_0,t_0)$  are arbitrary real constants.

We conclude that, if $Q^{(0)}(x,t)$, $W^{(0)}(x,t)$ and $\psi^{(0)}(x,t,k)$ are known, the
explicit expressions of $\rho(x,t)$ and $\hat{z}(x,t)$ given
above yield, via (\ref{realDDT}) and (\ref{diadicreal}), the DDT
matrix and therefore the new solution $Q(x,t)$, $W(x,t)$ through (\ref{realBacklund}).
As in the previous case, the corresponding formulas which give the
expression of the solutions $U(x,t)$, $W^{(\pm)}(x,t)$ of the equation
(\ref{redboomeron}) follow from the block structure of
$Q$ and $W$, of $\Sigma$, of the projector $\hat{P}$
\begin{equation}\label{realPblock}
\hat{P}=\left(\begin{array}{cc}\hat{B}^{(+)} & \hat{B}{}^\dagger \\
\hat{B} &
\hat{B}^{(-)}\end{array}\right)
\end{equation}
and of the vector $\hat{z}$
\begin{equation}\label{realzblock}
\hat{z}=\left(\begin{array}{c}\hat{z}^{(+)}  \\
\hat{z}^{(-)} \end{array}\right)\,.
\end{equation}
The relevant relations then read
\begin{subequations}\label{realUwpwmBacklund}
\begin{equation}\label{realUBacklund}
U=U^{(0)}+2\rho \hat{B}S^{(+)}\,,
\end{equation}
\begin{equation}\label{realwpBacklund}
W^{(+)}=W^{(0)(+)}+2\rho [C^{(1)(+)},\hat{B}^{(+)}S^{(+)}]\,,
\end{equation}
\begin{equation}\label{realwmBacklund}
W^{(-)}=W^{(0)(-)}+2\rho [C^{(1)(-)},\hat{B}^{(-)}S^{(-)}]\,,
\end{equation}
\end{subequations}
or, equivalently (see(\ref{diadicreal}), (\ref{realPblock}) and
(\ref{realzblock})) and more explicitly,
\begin{subequations}\label{realBacklundcompl}
\begin{equation}\label{realUBacklundcompl}
U=U^{(0)}+2\rho \frac{\hat{z}^{(-)}\hat{z}^{(+)}
{}^\dagger S^{(+)}}{<\hat{z}^{(+)},\hat{z}^{(+)}>+
<\hat{z}^{(-)},\hat{z}^{(-)}>}\,,
\end{equation}
\begin{equation}\label{realwpBacklundcompl}
W^{(+)}=W^{(0)(+)}+2\rho \frac{(C^{(1)(+)}\,\hat{z}^{(+)}\hat{z}^{(+)}{}^\dagger S^{(+)}-\hat{z}^{(+)}\hat{z}^{(+)}{}^\dagger S^{(+)}\,C^{(1)(+)})}{<\hat{z}^{(+)},\hat{z}^{(+)}>+
<\hat{z}^{(-)},\hat{z}^{(-)}>}\,,
\end{equation}
\begin{equation}\label{realwmBacklundcompl}
W^{(-)}=W^{(0)(-)}+2\rho \frac{(C^{(1)(-)}\,\hat{z}^{(-)}\hat{z}^{(-)}{}^\dagger S^{(-)}-\hat{z}^{(-)}\hat{z}^{(-)}{}^\dagger S^{(-)}\,C^{(1)(-)})}{<\hat{z}^{(+)},\hat{z}^{(+)}>+
<\hat{z}^{(-)},\hat{z}^{(-)}>}\,,
\end{equation}
\end{subequations}
where the expression of $\rho=\rho(x,t)$ is given by (\ref{roexpl}) with (see (\ref{functions}))
\begin{equation}\label{fghmcomp}
\begin{array}{l}
f(x,t)= <\hat{z}^{(+)},\hat{z}^{(+)}>+<\hat{z}^{(-)},\hat{z}^{(-)}>\,, \\
g(x,t)= <\hat{z}^{(+)},S^{(+)}\hat{z}^{(+)}>+<\hat{z}^{(-)},S^{(-)}\hat{z}^{(-)}>\,,     \\
h(x,t)= <\hat{z}^{(+)},U^{(0)}{}^\dagger S^{(-)}\hat{z}^{(-)}>-<\hat{z}^{(-)},S^{(-)} U^{(0)}\hat{z}^{(+)}>\,,\\
m(x,t)= <\hat{z}^{(+)},S^{(+)}C^{(1)(+)}\hat{z}^{(+)}>-<\hat{z}^{(-)},S^{(-)} C^{(1)(-)}\hat{z}^{(-)}>\,.      
\end{array}
\end{equation}

We end this section noticing that the explicit formulas derived
here and in the previous subsection \ref{sec:DDTcompl} are meant to serve as the main tools to construct soliton-- and, by
repeated application of DDTs, multisoliton--solutions of the matrix equations (\ref{redboomeron}). However, these formulas have been obtained by
algebra and local integration of differential equations. Therefore, the
two important properties of the solutions they yield, namely
their boundary values and their boundedness, are left out of our present discussion. These two issues, which are of basic relevance in applications, will be taken up in a following paper \cite{DL2} where explicit
expressions of solutions will be displayed and investigated.

\section{Conclusions and remarks}\label{sec:conclusions}
In this paper, we have set up the formalism of the Darboux--Dressing Transform (DDT) to construct a new solution of  the coupled matrix nonlinear integrable evolution PDEs 
(\ref{boomeron}),
\setlength\arraycolsep{2pt}
\begin{displaymath}
\begin{array}{lll}
Q_t &=&[C^{(0)},Q]+\sigma [C^{(1)},Q_x]-\sigma
\{Q,W\}-i\gamma \sigma(Q_{xx}-2Q^3)\,,\\
W_x &= &[C^{(1)},Q^2]\,, 
\end{array}
\end{displaymath}
from a given known solution of the same  equations. The choice of this system of  equations, which has been first introduced in \cite{CD2004}, is mainly motivated 
 by the fact that it captures several interesting models of dispersive propagation of multicomponent waves in weakly nonlinear media. 
In fact, it should be pointed out that setting $C^{(1)}=c\sigma$ implies that the linear terms
$[C^{(0)},Q]$ and $\sigma[C^{(1)},Q_x]$ can be both transformed away with the consequence that $W=0$ thereby obtaining the well known matrix NLS type equation
\begin{displaymath}
Q_t=-i\gamma \sigma (Q_{xx}-2Q^3)\,,
\end{displaymath}
which originates well known VNLS equations.
%Indeed, it is to point out that under appropriate conditions, it reduces to the well know VNLS (vector NLS) type equation (\ref{mother})
In contrast, whenever $C^{(1)}$ is a generic matrix, it is precisely the non commutative character of the matrix coefficients $C^{(0)}$ and $C^{(1)}$ which introduces the interesting phenomenology of boomerons, i.e. solitons which have different asymptotic velocities at $t=+\infty$ and $t=-\infty$. These special soliton solutions were first introduced long time ago in connection with the hierarchy of the matrix Korteweg--de Vries equation, but recently boomerons appeared again in geometry and optics and are attracting a growing interest in applications. % ??? 
Moreover and interestingly enough, the matrix evolution equation given above, which is the most general second order differential equation generated by the Lax pair (\ref{lax}), includes the  simpler class of evolution equations corresponding to the choice  $\gamma=0$ which turns out to be nonlinear and non trivial. This  class features first derivatives with respect to the space variable $x$ (rather than second derivatives as is the case of NLS type equations), and the simplest system of
equations of this type  coincides with
the standard equations describing the resonant interaction of 3 waves -- yet their solutions feature, somewhat unexpectedly, the boomeronic or trapponic behavior. This is  one instance of  wave equations which are both in our class and of well known applicative relevance. We trust the interested reader will be attracted by looking for other wave equations via reduction of our ``mother'' system  (\ref{boomeron}) 
either by exploring the examples of such equations reported in \cite{CD2004}, or by engaging himself/herself in deriving new equations.

The class of evolution equations we have chosen contains in fact, as reduced cases,  systems of wave equations which, because of their dispersion (mainly of Schr\"{o}dinger--type) and/or because of their nonlinearities, are, or promise to be, interesting models of applicative importance and it is with applications in mind that the DDT technique has been investigated and presented here almost as an algorithm to find explicit solutions. In its simplest version, this technique goes via a Darboux transformation of  the matrix solution of the Lax equations, which introduces just one pole in the complex plane of the spectral variable. Though this method is known, we show that new algorithmic features arise if the pole is on the real axis. This case, on the other hand, is the one which matters in the construction of solutions with non vanishing boundary values (as for dark solitons). The relation between the reality of the pole and the boundary values, together with explicit solutions, will be discussed in a separate paper.

%The relation between the reality of the pole and the boundary values is not discussed here as it will be the content of a separate paper.

\section*{Acknowledgments}
% \ack
The work of S L was partially supported by the EU GIFT project (NEST- Adventure Project no. 5006). Partial financial support has been provided also by the Italian ministry of education and research MIUR within the PRIN project SINTESI.

%%%%%%%%%%%%%%%%% Bibliography %%%%%%%%%%%%%%%%%%%

\end{document}